\documentclass[12pt]{article}
\usepackage{amssymb,graphicx}
\usepackage{epstopdf}
\usepackage{amsmath,amsfonts}
\usepackage{epsfig} 
\usepackage{graphicx,graphics}

\begin{document}
\title{\bf LHC signatures of vector boson emission from brane to bulk}
\author{D.~V.~Kirpichnikov$^{a,b}$\footnote{{\bf e-mail}: kirpich@ms2.inr.ac.ru}
 \\
$^a$ \small{\em Institute for Nuclear Research of the Russian Academy of Sciences} \\
\small{\em Prospect of the 60th Anniversary of October 7a, Moscow, Russia, 117312}\\
$^b$ \small{\em Moscow State University, Department of Physics,
}\\
\small {\em Vorobjevy Gory, 119991 Moscow, Russia
}
} 
\maketitle

\begin{abstract}
 In the framework of the RSII-$n$ model with $n$ compact and one
 infinite extra dimensions, we study the production of $Z$-bosons
 and photons, which escape into the bulk, in association with a jet 
 in $pp$ collisions at the LHC energies. This would show up as 
 the process $pp\to\mbox{jet}+\not\!\!\!E_T$. 
 We calculate the distributions in 
 the jet transverse momentum and rapidity and compare them with 
 the Standard Model background $pp\to \mbox{jet}+\nu\bar{\nu}$. 
 We find that the models with $n\ge 4$ can be probed at the  
 collision energy 14 TeV, while searches at 7 TeV are sensitive to 
 models with  $n\ge 6$ only.
\end{abstract}

\section{Introduction}

One potentially interesting property of the brane-world scenario 
with extra dimensions of infinite size is the emission of particles 
from the brane to the bulk 
 \cite{Rubakov:1983bb,Dubovsky:2000am,Gregory:2000rh}.
 This is one of the ways in which extra dimensions may open up 
 in experiment
\cite{Gninenko:2003nx,Friedland:2007yj,Gninenko:2008yq,Astakhov:2010ji}
 and also in astroparticle 
physics and cosmology
\cite{Friedland:2008zza,MosqueraCuesta:2002bx}. A  four-dimensional 
observer residing on the brane is unable to detect particles 
escaping to infinity in extra dimensions, so the observable 
signature is missing energy. This is analogous to the ADD model \cite{ADD}, in which
the Kaluza-Klein gravitons are also undetectable, and their 
emission would also manifest itself as missing energy \cite{GRW}.
This analogy goes further: in the case of infinite extra 
dimensions, the spectrum of four-dimensional masses of escaping 
particles is continuous, while the KK graviton spectrum
is also nearly  continuous in the ADD model.

A particular setup that ensures the gravitational quasi-localization
of various bulk fields on the brane is a model with one infinite 
and several compact extra dimensions, equipped with the AdS metric
\cite{Gregory,Gherghetta:2000qi,Oda,Dubovsky:2000av,Dubovsky:2002xv}. 
This is a mild 
generalization of the original Randall-Sundrum II model 
\cite{Randall:1999vf}. Interestingly, unlike the RSII geometry,
the generalized setup quasi-localizes bulk gauge fields 
\cite{Oda,Dubovsky:2002xv}. So, it makes 
sense to consider the emission
of vector particles from the brane to the bulk and employ this setup as a concrete example.

In our previous paper \cite{Astakhov:2010ji} we performed
a phenomenological analysis of the vector boson emission to the bulk 
in $e^+e^-$  collisions. In that case, the promising process is
$e^+e^-\to\gamma+\not\!\!E_T$. In this paper we extend the 
analysis to $pp$ collisions at the LHC energies and study the process 
$pp \to \mbox{jet}+\not\!\!\!E_T$, where energy is carried away by 
either photon or $Z$-boson emitted into the bulk. Important 
constraints on the parameters of the setup come from the analysis of the 
invisible  $Z$-boson decay \cite{Gninenko:2008yq} which is due to 
the fact that the massive $Z$-boson is quasi-localized, rather 
then exactly localized on the brane. With these constraints, the 
prospects of observation of the process we study are not particulary
bright for $pp$ collisions at 7 TeV: its cross section is small 
compared to the Standard Model (SM) background $pp\to \mbox{jet}+\nu\bar{\nu}$
unless the number of compact extra dimensions is large, $n\ge 6 $.
The situation is better for $pp$-collision energy of 14 TeV. In this case, models with smaller numbers of compact extra dimensions
can be probed.

This paper is organized as follows. 
In Sec.~\ref{section3}  we consider the   Standard Model in the 
background  of warped $(4+1+n)$-dimensional mectric. We put 
$SU(2)_L\times U(1)_Y$  gauge
sector of   SM into the bulk, while, for definitness,  the  SM 
fermions are supposed 
to be  localized  on the brane. In Sec.~\ref{section5}  we consider bulk dynamics of the 
gauge  fields. 
In Sec.~\ref{section6} we derive  differential
rates of the processes  $pp \to \mbox{jet}+Z_{bulk}$, $pp \to \mbox{jet}+\gamma_{bulk}$.
 In  Sec.~\ref{section7} we present our results and compare  the rates with the background process $pp \to \mbox{jet} +\nu\bar{\nu}$. 
 We conclude in Sec.~\ref{summary}.
 
\section{$SU(2)_L\times U(1)_Y$ bulk sector of the Standard Model\label{section3}}
Let us consider $AdS_{n+5}$ metric with
$n$  extra dimensions compactified on a torus $T^{n}$ and one infinite extra dimension,
\begin{equation}
ds^2=a^2(z)(\eta_{\mu \nu} dx^\mu dx^\nu -\delta_{ij} d\theta^i d\theta^j ) - dz^2 =G_{MN} dx^M dx^N, \label{metrics of the theory}
\end{equation}
where   $M,N =  0, 1, 2, 3, 5,...,n+4, n+5$,  
indices $\mu, \nu $ run from $0$ to $3$, indices $i,j$ run from $5$ 
to $(n+4)$ and refer to compact extra dimensions, $\theta^i \in [0,2\pi 
R_i]$,  $R_i$ are the radii of the compact exrta dimensions,
 the coordinate 
$x^{n+5}\equiv z$ denotes a non-compact spatial dimension. The
warp factor is 
 $$a(z)=e^{-k|z|}.$$
  The metric 
(\ref{metrics of the theory}) is a solution to the $(n+5)$-dimensional 
Einstein equations with fine-tuned bulk cosmological constant and 
brane tension. The parameter $k$ is determined by the $(5+n)$-
dimensional Planck mass and bulk cosmological constant;
it is the only free parameter of the model. We call this 
setup RSII-$n$ model.

We consider $(5+n)$-dimensional $SU(2)_{L}\times
U(1)_Y$ gauge theory with bulk gauge fields $A^\alpha_{M}$, $B_M$ 
and scalar doublet
$\Phi$ in the background metric (\ref{metrics of the theory}). 
We assume for definiteness that fermions $\psi$ are localized on the 
brane and hence depend only on four-dimensional coordinates $x$. The 
action of this model  is
\begin{equation}
S=\int d^4 x\,dz\prod^n_{i=1} \frac{d\theta_i}{2\pi
R_i}\sqrt{g}\left[-\frac{1}{4}\left(F^\alpha_{MN}\right)^2
-\frac{1}{4}B^2_{MN}+\left(D_M \Phi\right)^\dagger D_M\Phi
-V(\Phi
)+\delta(z)\mathcal{L}_F\right],\label{theactionofthetheory}
\end{equation}
where $\mathcal{L}_F$ is the fermion Lagrangian,
$$
\mathcal{L}_{F}=
i\bar{q}_L\left(\hat{\partial}-i\widetilde{g}_{1}\frac{Y^q_{L}}
{2}\hat{B}-i\widetilde{g}_{2}\frac{\sigma_{i}}{2}A^{\alpha}\right)q_L
 +i\bar{q}_R\left(\hat{\partial}-i\widetilde{g}_{1}\frac{Y^{q}_{R}}
 {2}\hat{B}\right)q_R,$$
 where we consider light quarks only and 
neglect their masses; summation over quark flavors is assumed. Here 
$\widetilde{g}_2$ and $\widetilde{g}_1$ 
are the
$SU(2)_L\times U(1)_Y$ bulk couplings, respectively, and
$V(\Phi)$ is the standard Higgs potential. We consider
the Higgs phase of the theory and perform the usual redefinition of 
the  gauge  fields,
$$Z_{M}=\frac{1}{\sqrt{\widetilde{g}^2_{1}+\widetilde{g}^2_{2}}}\left(-\widetilde{g}_{1}B_{M}+\widetilde{g}_{2}A^3_{M}\right),
\quad
A_{M}=\frac{1}{\sqrt{\widetilde{g}^2_{1}+\widetilde{g}^2_{2}}}\left(\widetilde{g}_{2}
B_{M}+\widetilde{g}_{1}A^3_{M}\right),
$$
$$W_{M}^\pm=\frac{1}{\sqrt{2}}\left(A_{M}^{1}\mp
iA_{M}^{2}\right).$$ Then
 the relevant part of the action (\ref{theactionofthetheory}) takes 
 the following form:
$$
S=\int d^4 x\,dz\prod^n_{i=1} \frac{d\theta_i}{2\pi R_i}\sqrt{g}
\Bigl[-\frac{1}{4}F^2_{MN}-\frac{1}{2}|\,W_{MN}\,|^2+m^2_W
|\,W_M\,|^2-\frac{1}{4}Z^2_{MN}+\frac{1}{2}m^2_Z
Z^2_M
 + \delta(z)\mathcal{L}_{f}\Bigr], 
 $$
 where $m_W^2=\frac{1}{4}\widetilde{g}_{2}^2 v^2$ and
$m_Z^2=\frac{1}{4}(\widetilde{g}_{2}^2+\widetilde{g}_{1}^2)v^2$
are the bulk masses squared of the gauge fields.
Note that photon remains  massless. Here $\mathcal{L}_f$ is the quark
Lagrangian in the Higgs phase,
$\mathcal{L}_f=\mathcal{L}_{f,EM}+\mathcal{L}_{f,W}$,
where
$$\mathcal{L}_{f,EM}=e_{(5)}\sum_q Q_q \,\bar{q}\gamma^\mu A_\mu q,$$
$$
\mathcal{L}_{f,W} = \frac{\widetilde{g}_2}{2\sqrt{2}} \left( 
\bar{u} \gamma^\mu (1-\gamma^5) W^+_\mu  d +h.c.\right)
+\frac{\widetilde{g}_2}{2 \cos \theta_W} \sum_{q=u,d} \bar{q} 
\gamma^\mu \left( T^3_q (1-\gamma^5) - 2Q_q \sin^2 \theta_W\right)   
Z_\mu q,$$
where $e_{(5)}= \widetilde{g}_2 \sin\theta_W =\widetilde{g}_1
\cos\theta_W $ is the bulk electromagnetic coupling  and $\,Q_q$ is the quark electric charge. 

\section{Bulk dynamics of the gauge fields \label{section5}}
\subsection{ Bulk Z-boson \label{bulkZ}}
In this section we study the properties of the bulk vector fields.
We assume that the sizes of compact extra dimensions are small, so that
at energies of interest we have $\sqrt{s} \ll 1/R_i$. Then the masses 
of KK modes inhomogeneous in compact dimensions are large and we can 
ignore KK excitations on the torus. Hence, we take into account only
the gapless and continuous KK spectrum corresponding to the motion along the $z$ 
direction. Before proceeding further, let us discuss the properties 
of the usual $Z$-boson in this model. Since the fermions couple only
to the components $Z_\mu$ of the field $Z_M$, it is consistent
to set $Z_5=...=Z_{n+5}=0$ and split the $Z$-boson wave function 
into  longitudinal  and transverse parts, 
$Z^\mu=p^\mu Z_L(p,z)+\epsilon^\mu  Z_T(p,z)$. Once the fermion masses
are neglected, the fermion currents are conserved on the brane,
so the longitudinal $Z$-bosons are  not emitted.   The equation of 
motion for the transverse mode 
$Z_T(p,z)$ is
\begin{equation}
\left(-\partial^2_5 +(2+n) k \, \mbox{sign}(z)\partial_5+ m^2_Z -
\frac{m^2}{a^2}\right)Z_m(z)=0,  \label{Z equation
tr}
\end{equation}
where $Z_T(m,z)\equiv Z_m(z) $ and $m$ is the four-dimensional
mass. We also note that odd eigenmodes of (\ref{Z equation
tr}) do not interact with the fermions localized on the brane.  
Even eigenmodes $Z_m(z)$ are normalized with the measure
$e^{-nk|z|}$:
\begin{equation}
\int dz \, e^{-nk|z|} Z_m(z)Z_{m'}(z)=\delta(m-m'). \label{norm_cond_gapless}
\end{equation}
\begin{figure}[t]
\begin{center}
\includegraphics[width=0.9\textwidth]{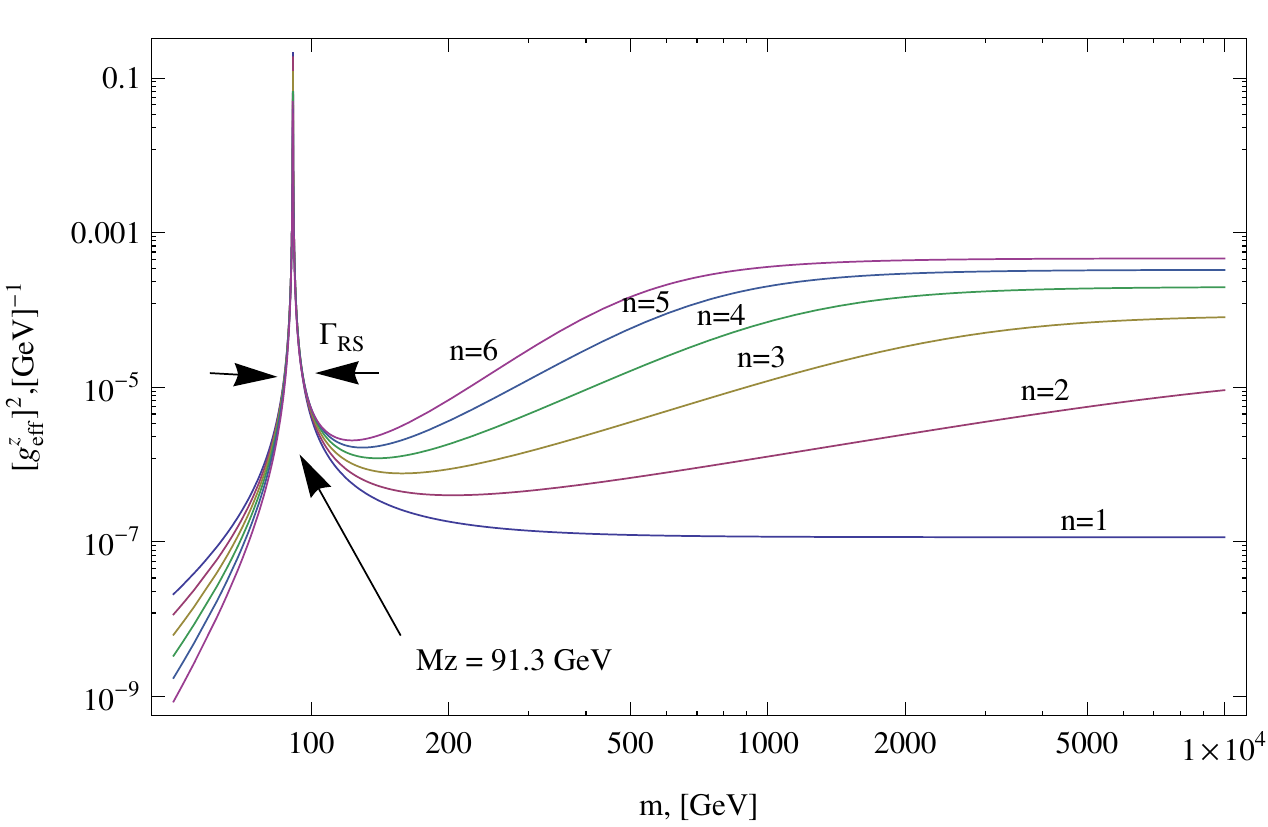}
\end{center}
{\caption \qquad { \em Bulk Z-boson  coupling parameter squared 
 versus four-dimensional mass $m$ for various 
numbers of extra dimensions   $n$. The values of $k$ for different 
$n$ are given in Table~\ref{table1}. } \label{figure2} }
\end{figure}
 The solution to Eq. (\ref{Z equation tr}) is
\begin{equation}
Z_m(z)=\sqrt{\frac{m}{2k}} \,e^{(\frac{n}{2}+1)k|z|}\left[a_m
J_\nu\left(\frac{m}{k}e^{k|z|}\right)+b_m
N_\nu\left(\frac{m}{k}e^{k|z|}\right) \right], \label{Z_m(z)}
\end{equation}
where the order of the Bessel functions is 
$\nu=\sqrt{\left(1+n/2\right)^2+(m_Z/k)^2}$.
The coefficients $a_m$ and $b_m$ are determined by the boundary condition on the brane
$\partial_z Z_m(+0)-\partial_z Z_m(-0)=0$ and the
normalization condition $a^2_m+b^2_m=1$, following from (\ref{norm_cond_gapless}).  We find 
$$
a_m=-\frac{B_m}{\sqrt{B^2_m+1}}, \,\,
b_m=\frac{1}{\sqrt{B^2_m+1}},$$ where  
\begin{equation}
B_m=\frac{N_{\nu-1}(\frac{m}{k})+\frac{k}{m}(\frac{n}{2}+1-\nu)N_\nu(\frac{m}{k})}
{J_{\nu-1}(\frac{m}{k})+\frac{k}{m}(\frac{n}{2}+1-\nu)J_\nu(\frac{m}{k})}.\label{coefficients_2}
\end{equation}
In the low energy limit only modes with $m \ll k$  are
relevant, and the expression (\ref{coefficients_2}) simplifies,
$$
B_m=-\frac{2\,\Gamma^2(\frac{n}{2}+1)}{\pi n} \cdot
\left(\frac{m}{2k}\right)^{-n}\left(
1-\frac{m^2_Z}{m^2}\frac{n}{(n+2)}\right).
$$
In this limit, the squared  wave function  on the brane is
$$
Z^2_m(0)=\frac{\Gamma^2(\frac{n}{2}+1)}{\pi^2(1+B^2_m)}\cdot\left(\frac{m}{2k}\right)^{-n-1}=\frac{n^2}{4}\cdot
\frac{m^4 }{ \Gamma^2(\frac{n}{2}) }\cdot \left(
\frac{m}{2k}\right)^{n-1} \frac{1}{(m^2-M^2_Z)^2+m^2
\Gamma_{RS}^{2}(m)},
$$
where
\begin{equation}\Gamma_{RS}(m)=\frac{2\pi}{n\Gamma^2\left(\frac{n}{2}\right)} m
\left(\frac{m}{2k}\right)^n, \quad M_Z=m_Z\sqrt{\frac{n}{n+2}}. \label{invisible_width}
\end{equation}
Hence, for $k\gg M_Z$, $Z$-boson is quasi-localized
on the brane, and $M_Z$ and $\Gamma_{RS}(M_Z)$ are its mass and
invisible decay width, respectively. In particular,    
 $ Z^2_m(0)$ 
tends to the delta function as $ \Gamma_{RS}\rightarrow 0$:
\begin{equation}
Z^2_m(0)=\frac{nk}{2}\cdot \frac{1}{\pi}
\frac{\Gamma_{RS}}{2}\frac{1}{(m-M_Z)^2+\left(\frac{\Gamma_{RS}}{2}\right)^2}\rightarrow
\frac{nk}{2} \cdot \delta(m-M_Z). \label{deltafunc}
\end{equation}
This yields the relation between the four-dimensional 
and five-dimensional couplings, $e_{(4)} = e_{(5)}\sqrt{\frac{nk}{2}}$.

The fact that $Z$-boson is not exactly localized on the brane implies
the lower bounds  on the parameter $k$. They come from the requirement
that the invisible decay width of $Z$-boson does not exceed the 
experimental uncertainty~\cite{Amsler}:
$$\Gamma_{RS}(M_Z)\le \Delta\Gamma^Z_{tot}=1.5 \,\mbox{MeV}.$$
These bounds are collected in Table~\ref{table1}. When presenting
numerical results, we will consider the values of $k$ saturating these 
bounds.
\begin{center}
    \begin{tabular}{|c |c |c |c |c |c |c | }
  \hline
    $n$   & $1$ & $2$ & $3$ & $4$ & $5$ & $6$ \\
    \hline
    $k , \mbox{GeV}$ & $5.5 \cdot 10^6$ & $2 \cdot 10^4$& $2.5 \cdot 10^3$ &$900$ & $400$  & $300$ \\
    \hline  
    \end{tabular}
\begin{center} 
      { \tablename\, \ref{table1}: \label{table1} {\em \,\, The  lower bounds on the parameter 
   $k$ for various numbers of compact extra dimensions $n$. }  }  
\end{center}
 \end{center} 
Coming back to the general discussion, we collect (\ref{Z_m(z)}) and 
(\ref{coefficients_2}) and find the expression for the wave
 function at the brane:
\begin{equation}
Z_m(0)=\frac{1}{\pi} \sqrt{\frac{2k}{m}}\frac{1}{\sqrt{\left[N_{\nu-1}
(\frac{m}{k})+\frac{k}{m}(\frac{n}{2}+1-\nu)N_\nu(\frac{m}{k})\right]^2+
\left[J_{\nu-1}(\frac{m}{k})+\frac{k}{m}(\frac{n}{2}+1-\nu)J_\nu(\frac{m}{k})\right]^2}}. \label{Z_m(0)tot}
\end{equation}
This wave function determines the coupling of fermions to the 
mode $Z_m$. Hence, it is useful to introduce the $Z$-boson coupling
parameter  ${\tt g^Z_{eff}}(m)=\sqrt{\frac{2}{nk}}Z_m(0)$ . 
Expanding the Bessel  functions at large argument in 
 (\ref{Z_m(0)tot}) one finds ${\tt g^Z_{eff}}(m)\simeq\sqrt{\frac{2}
 {\pi nk}}$ 
for $m/k\gg 1$.  We show $[{\tt g^Z_{eff}}(m)]^2$ as function of $m$ 
in Fig.~\ref{figure2}. Away from the $Z$-pole, the effective coupling 
increases with $m$ and flattens out at large $m$. Note that the 
curves in Fig.~\ref{figure2} correspond to different values of $k$.
This explains the fact that the large-$m$ asymptotics in 
Fig.~\ref{figure2} are higher for larger $n$, while for fixed $k$,
the asymptotic values of the effective coupling decrease with $n$
as ${\tt g^Z_{eff}}\sim n^{-1/2}$.

\subsection{Bulk photon}
Let us now turn to the bulk photon.
We again set $A_5=...=A_{n+5}=0$.
 The equations of motion for the bulk photon are
\begin{equation}
p^\mu \partial_z A_\mu=0,
\label{eq1xphoton moment}
\end{equation}
\begin{equation}
\left(-\partial^2_z +(2+n) k \, \mbox{sign}(z)\partial_z-
\frac{p^2}{a^2} \right)A_\lambda +\frac{1}{a^2}p^\mu
p_\lambda A_\mu =0, \label{eq2xphoton moment}
\end{equation}
 Eqs. (\ref{eq1xphoton moment}) and (\ref{eq2xphoton moment})  
have a constant solution with respect to the $z$-coordinate, 
$A^{(0)}_\mu(p,z)\equiv A^{(0)}_\mu(p)=\mbox{const}$.
 This zero mode describes photon localized on
the brane. The normalization condition is
\begin{equation}
\int^\infty_\infty dz\, e^{-nk|z|}\, |A^{(0)}|^{\,2}=1, 
\end{equation}
which gives $A^{(0)}=\sqrt{\frac{nk}{2}}$.
There is also gapless continuum of bulk modes:
\begin{equation}
A_m(z)=\sqrt{\frac{m}{2k}} \,e^{(\frac{n}{2}+1)k|z|}\left[f_m
J_{\frac{n}{2}+1}\left(\frac{m}{k}e^{k|z|}\right)+g_m
N_{\frac{n}{2}+1}\left(\frac{m}{k}e^{k|z|}\right) \right],
\label{bulk photon}
\end{equation}
with the normalization condition
$f^2_m+g^2_m=1$
and boundary condition on the brane
$ \partial_z A_m(+0)-\partial_z A_m(-0)=0$.
Explicitly, Eq. (\ref{bulk photon}) reads:
\begin{figure}[t]
\begin{center}
\includegraphics[width=0.9\textwidth]{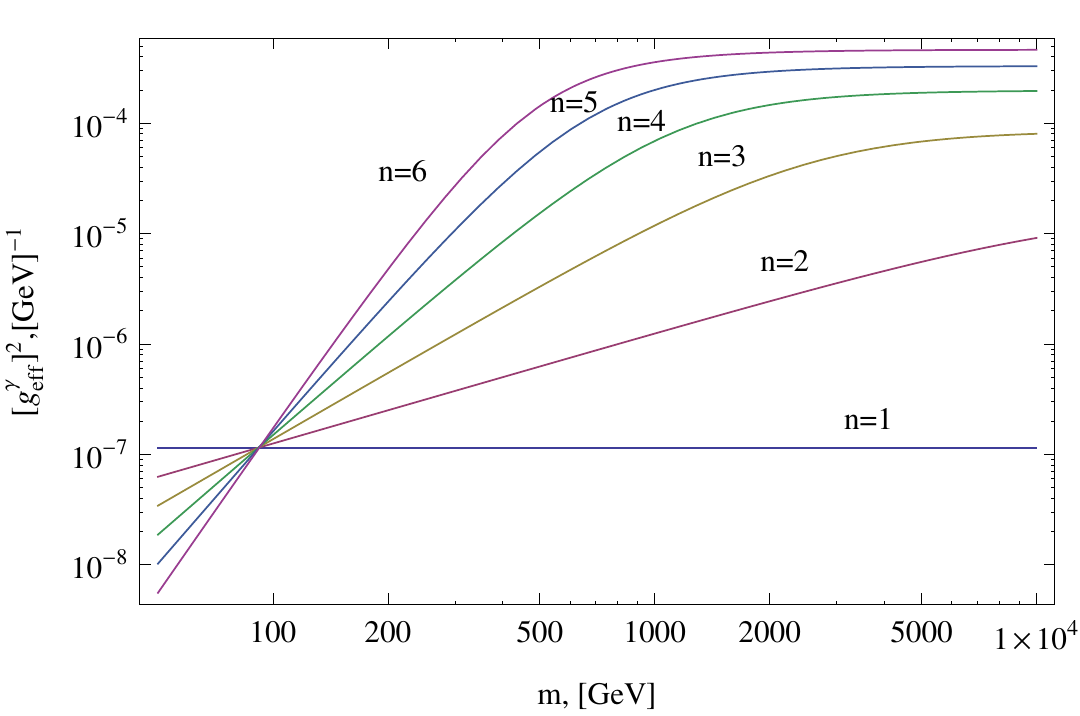}
\end{center}
{\caption \qquad { \em Bulk  photon coupling squared versus 
four-dimensional mass $m$ for various  numbers of extra dimensions   
$n$.  The values of $k$ for different 
$n$ are given in Table \ref{table1}. } \label{figure3} }
\end{figure}  
\begin{equation}
A_m(z)=\sqrt{\frac{m}{2k}}
\,e^{(\frac{n}{2}+1)k|z|}\frac{\left[N_{\frac{n}{2}}\left(\frac{m}{k}\right)
J_{\frac{n}{2}+1}\left(\frac{m}{k}e^{k|z|}\right)-J_{\frac{n}{2}}\left(\frac{m}{k}\right)
N_{\frac{n}{2}+1}\left(\frac{m}{k}e^{k|z|}\right)
\right]}{\sqrt{J^2_{\frac{n}{2}}\left(\frac{m}{k}\right)+N^2_{\frac{n}{2}}\left(\frac{m}{k}\right)}},
\label{bulk photon final}
\end{equation}
The interaction of these modes with the brane fermions is determined 
by their wave functions at the brane, which are given by
\begin{equation}
A_m(0)=\frac{1}{\pi} \sqrt{\frac{2k}{m}}\frac{1}{\sqrt{N^2_{\nu-1}
(\frac{m}{k})+
J^2_{\nu-1}(\frac{m}{k})}}. \label{photon_WF_at_zero}
\end{equation}
Like in Section \ref{bulkZ}, we introduce the coupling parameter 
${\tt g^\gamma_{eff}}(m)=\sqrt{\frac{2}{nk}}A_m(0)$. For the 
relatively light modes with $m/k \ll 1$ it is suppressed, 
${\tt g^\gamma_{eff}}(m)\simeq \sqrt{\frac{n}{2k}} \frac{1}{ 
\Gamma(\frac{n}{2})} \left( \frac{m}{2k}\right)^{\frac{n-1}{2}}$.
However, this suppression disappears at high energies, 
${\tt g^\gamma_{eff}}
(m)\simeq \sqrt{\frac{2}{\pi n k}}$ at $m/k\gg 1$. In  Fig. 
\ref{figure3} we show 
$[{\tt g^\gamma_{eff}}(m)]^2$ for various numbers of extra 
dimensions $n$. We again note that the curves in Fig.~\ref{figure3} 
correspond to different values of $k$.

Two remarks are in order. First, models with gapless spectra 
of photons, like the one considered in this paper, are strongly 
constrained by low energy physics experiment
\cite{Gninenko:2003nx,Friedland:2007yj}
and astrophysics \cite{Friedland:2007yj,Friedland:2008zza}. We are not going to use
these constraints in what follows, since they can be evaded by 
giving a relatively small gap to the bulk vector bosons (see Ref. 
\cite{Smolyakov:2011hv} for concrete example). Second, interactions of the 
photon zero mode with bulk fields is potentially dangerous
\cite{Smolyakov:2011hv}, since this mode is inhomogeneous in the extra 
dinmension $z$. Likewise, the interaction of the
quasi-localized $Z$-boson with bulk fields is potentially
dangerous, so having $SU(2)_L$ gauge theory in the bulk may be
problematic. We leave this issue for further analysis and proceed
with our phenomenological study at the linearized level.
  
\section{The process $pp\to \mbox{jet}+Z_{bulk}(\gamma_{bulk})$ \label{section6}}
\begin{figure}[t]
\begin{center}
\includegraphics[width=0.9\textwidth]{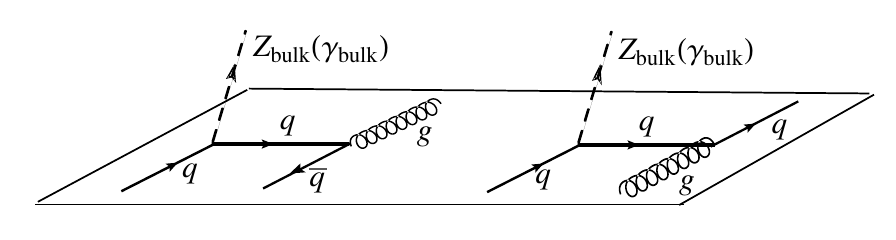}
\end{center}
{\caption \qquad{\em The process $pp\to \mbox{jet}++Z_{bulk}
(\gamma_{bulk})$. }
 \label{diagrams} }
\end{figure}  
Within the RSII-$n$ model, the most promising process to 
search for at $pp$ colliders is 
$pp \to \mbox{jet}+Z_{bulk}(\gamma_{bulk})$,
see Fig. \ref{diagrams}, where the jet originates from gluon 
or quark, and $Z_{bulk}$ and $\gamma_{bulk}$ manifest themselves
as missing energy. In this Section we derive the 
rate of this process. We express it  in terms of
differential cross section, where the contributions of different
KK modes of both bulk $Z$-boson and bulk photon have been summed up. In 
the RSII-$n$ model this sum 
is actually the integral over $m$. The  differential  cross  
sections  for the   parton subprocesses 
$\bar{q}q \to g Z_{bulk}(\gamma_{bulk})$, 
$ g\bar{q} \to \bar{q} Z_{bulk}
(\gamma_{bulk})$ and
$gq\to q Z_{bulk}(\gamma_{bulk})$ are written as follows:
\begin{equation}
d\sigma^{ij}=(2\pi)^4 \delta^{(4)}(p_i+p_j-p_k-q)
\frac{\overline{\sum}|\mathcal{M}^{ij}|^2}{4 N_{ij}I }\, 
\frac{d^3 p_k}{(2\pi)^3 2p^0_k}\frac{d^3 q}{(2\pi)^3 2q^0}\,dm,
 \label{dif_parton_cross_sect1}
\end{equation}
\begin{figure}[t]
\begin{center}
\includegraphics[width=0.9\textwidth]{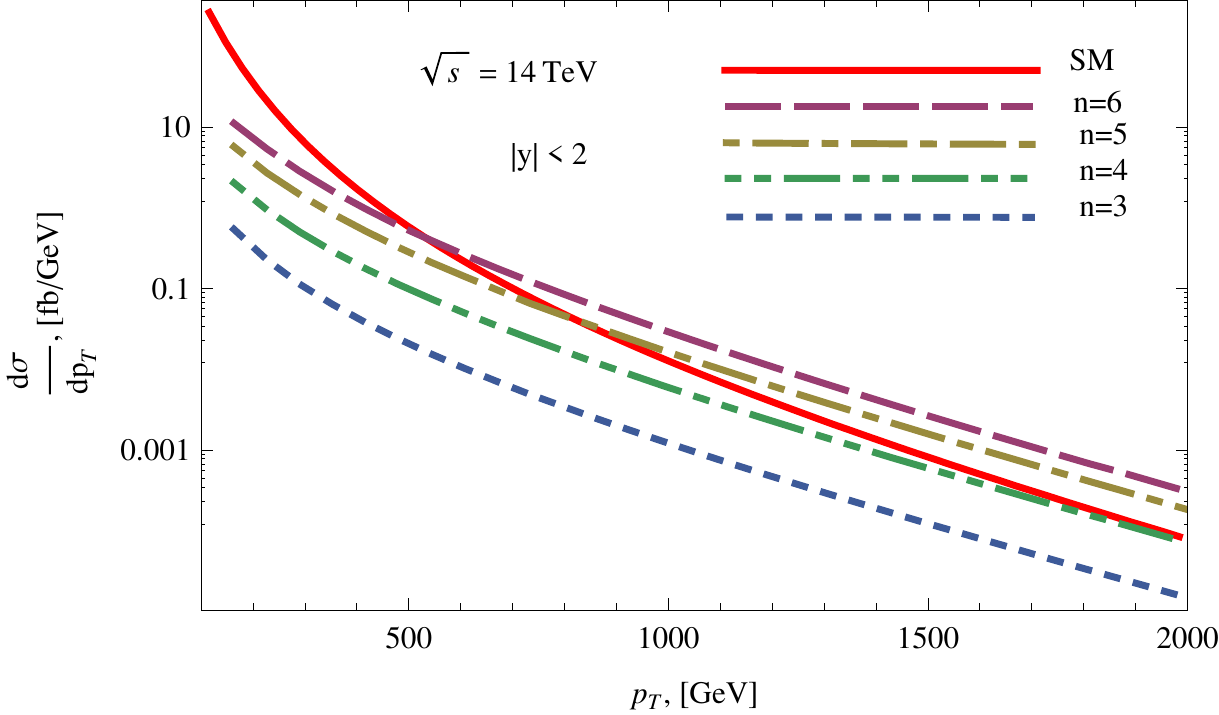}
\end{center}
{\caption \qquad {\em Differential cross section of the process 
$pp \to \mbox{jet}+Z_{bulk}(\gamma_{bulk})$ (dashed lines)  versus the jet 
transverse momentum $p_T$ for
various numbers of compact extra dimensions $n=3,4,5,6$. The 
rapidity of a jet is integrated within the interval $|y|< 2$. The 
Standard Model background is  $pp \to \mbox{jet}+\nu\bar{\nu}$ (solid line). 
The center-of-mass energy of incoming protons is $\sqrt{s}=14$TeV. The 
values  of $k$  are given in Table \ref{table1}}. 
 \label{figure1}}
\end{figure}  
here $I=(p_i p_j)$, 
the partons are treated as massless, $m$  
is the four-dimensional mass of $Z$-boson or photon, whose
dispersion relation is $m^2=q^2$, indices $i,j$ 
denote the incoming, and $k$ outgoing parton states $(q,\bar{q},g)$. 
The sum  $$\overline{\sum}=\frac{1}{4}\sum_{pol}\sum_{col}$$
runs over polarization and color. The factor 
$N_{ij}$ comes from the parton color averaging, it is equal to 
$N_{\bar{q}q}=N_c^2$ and
 $N_{\bar{q}g}=N_{qg}=N_c(N_c^2-1)$. 
The energies of outgoing parton and bulk particles are equal to
$p^0_k=|{\bf p}_k|$ and $q^0=\sqrt{m^2+{\bf q}^2}$, respectively.
The four-momenta of incoming partons are
$p_i=(x_1 \sqrt{s}/2,0,0,x_1\sqrt{s}/2), \quad p_j=(x_2 \sqrt{s}/2,0,0,-
x_2\sqrt{s}/2),$
where $\sqrt{s}$ is the collider center-of-mass energy.
In the following calculation we denote $p_3=p_{i3}+p_{j3}$, 
 $p_T=\sqrt{p^2_{k1}+p^2_{k2}}$ and $p_{k3}=p_T \sinh y$, where 
 $y$ is the rapidity of the outgoing parton.   
The energies of the outgoing  particles  can be rewritten as 
$p^0_k=\sqrt{p^2_T+p_{k3}^2}=p_T \cosh y$ and 
$q^0=\sqrt{m^2+p^2_T+(p_3-p_T \sinh y)^2}$.
\begin{figure}[t]
\begin{center}
\includegraphics[width=0.9\textwidth]{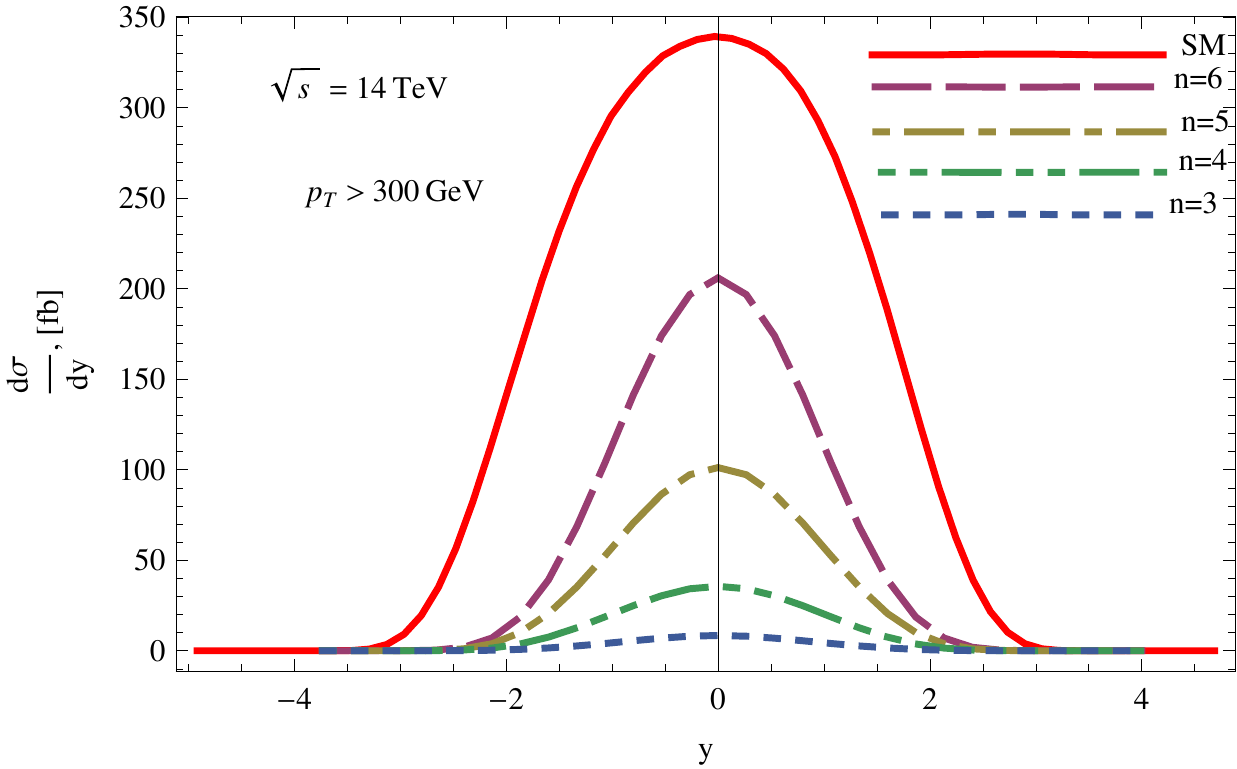}
\end{center}
{\caption \qquad{\em Differential cross section of the process 
$pp \to \mbox{jet}+Z_{bulk}(\gamma_{bulk})$ (dashed lines)  versus the jet 
rapidity $y$  for
various numbers of compact extra dimensions $n=3,4,5,6$. The  jet
transverse momentum is integrated over the range $p_T>300$ GeV. The 
Standard Model background is  $pp \to \mbox{jet}+\nu\bar{\nu}$ (solid line). 
The center-of-mass energy of incoming protons is $\sqrt{s}=14$TeV. The 
values  of $k$  are given in Table \ref{table1}.}
 \label{figure4} }
\end{figure}  
 The Mandelstam variables are equal to
$$\hat{s}=(p_i+p_j)^2=x_1x_2 s,
 \quad
 \hat{t}=(p_i-p_k)^2=-x_1 p_T \sqrt{s}\, e^{-y}, \quad 
\hat{u}=(p_j-p_k)^2=-x_2 p_T \sqrt{s}\, e^{y}.$$

The relation 
$\hat{s}+\hat{t}+\hat{u}=m^2$ gives
\begin{equation}
m^2=x_1x_2 s-x_1 p_T \sqrt{s}\, e^{-y}-x_2 p_T \sqrt{s}\, e^{y}=
x_1x_2 s\left(1-\frac{1}{2}\frac{x_T}{x_2}e^{-y}-
\frac{1}{2}\frac{x_T}{x_1}e^{y}\right)\ge 0, \label{kinematical_area}
\end{equation}
where  $x_T=2p_T /\sqrt{s}$ is the fractional transverse 
energy of the outgoing parton.
 The  inequality (\ref{kinematical_area}) defines the kinematically 
 allowed region for the subprocesses with particles escaping from our 
 brane. 
The differential cross section of the process $pp \to \mbox{jet}+Z_{bulk}(\gamma_{bulk})$ is written as follows:
$$
\frac{d\sigma}{dp_{T}dy}(pp \to \mbox{jet}+Z_{bulk}(\gamma_{bulk}))=\sum_{q=u,d} \int^1_0 dx_1\int^1_0 dx_2 \,\,
\theta(m^2)  $$
$$\times \Bigl\{\Bigl[ f_q(x_1,\mu)f_{\bar{q}}(x_2,\mu)+f_q(x_2,\mu)f_{\bar{q}}(x_1,\mu)\Bigr] \frac{d^2 \sigma}{ dp_T dy}(\bar{q}q\to g Z_{bulk}(\gamma_{bulk}))  $$
$$
+\Bigl(\Bigl[ f_q(x_1,\mu)f_{g}(x_2,\mu)+f_{\bar{q}}(x_1,\mu)f_{g}(x_2,\mu)\Bigr] \frac{d^2 \sigma}{ dp_T dy}(qg\to q Z_{bulk}(\gamma_{bulk})) 
$$
\begin{equation} +\Bigl[f_q(x_2,\mu)f_{g}(x_1,\mu)+f_{\bar{q}}(x_2)f_{g}(x_1,\mu)\Bigr]\frac{d^2 \sigma}{ dp_T dy}(gq\to q Z_{bulk}(\gamma_{bulk}))  \Bigr) \Bigr\} \label{DifCrSctZgmm}
\end{equation}
where $f_i(x,\mu)$ are parton distribution functions.
The parton differential rates are obtained by integrating
Eq.~(\ref{dif_parton_cross_sect1}) over $m$ and ${\bf q}$:
 \begin{equation}
 \frac{d\sigma}{  dp_T \, dy}(ij\to k Z_{bulk}(\gamma_{bulk})) = \frac{p_T}{8\pi N_{ij} m\, x_1 x_2 s} \overline{\sum}|\mathcal{M} (ij\to k Z_{bulk}(\gamma_{bulk}))|^2. 
 \end{equation}
 \begin{figure}[t]
\begin{center}
\includegraphics[width=0.9\textwidth]{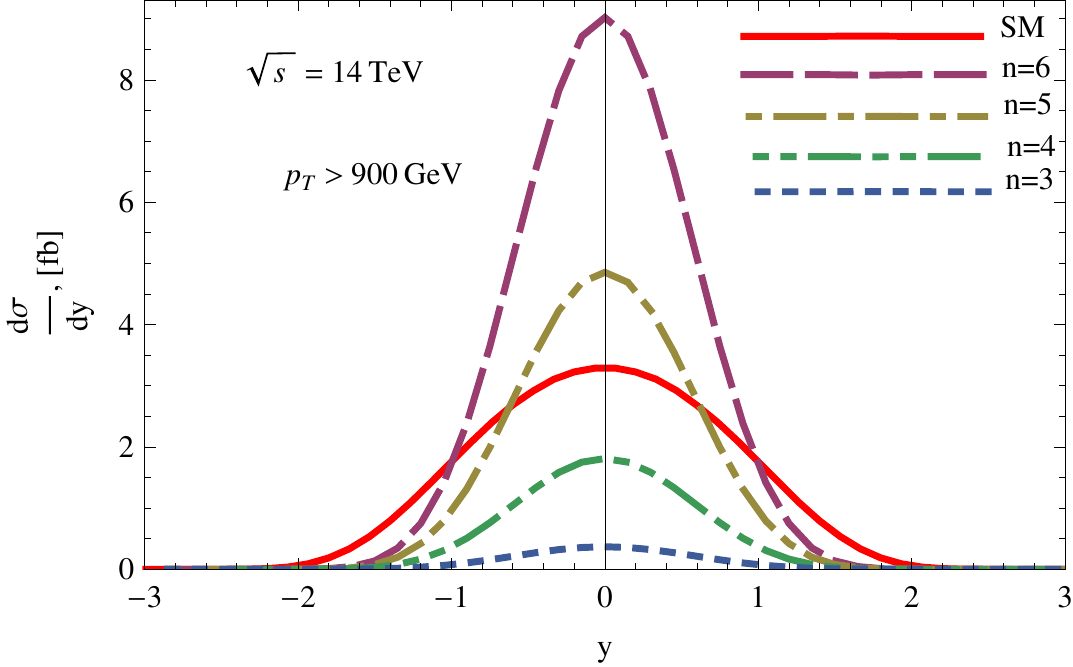}
\end{center}
{\caption \qquad{\em  Same as in Fig. \ref{figure4}, 
but with the transverse momentum integrated over the
 range $p_T>900$ GeV.}  \label{figure5} }
\end{figure}
Squared matrix elements for the subprocesses  obey the crossing symmetry relations
\begin{equation}
\overline{\sum}|\mathcal{M}(gq\to qZ_{bulk}(\gamma_{bulk}))|^2
=-\overline{\sum}|\mathcal{M}(\bar{q}q\to gZ_{bulk}(\gamma_{bulk}))|^2\Big{|}_{\hat{s}\leftrightarrow\hat{t}},\label{cross_sym1}
\end{equation}
\begin{equation}
\overline{\sum}|\mathcal{M}(qg\to qZ_{bulk}(\gamma_{bulk}))|^2
=-\overline{\sum}|\mathcal{M}(\bar{q}q\to gZ_{bulk}(\gamma_{bulk}))|^2\Big{|}_{\hat{s}\leftrightarrow\hat{u}}. \label{cross_sym2}
\end{equation}
\begin{figure}[t]
\begin{center}
\includegraphics[width=0.9\textwidth]{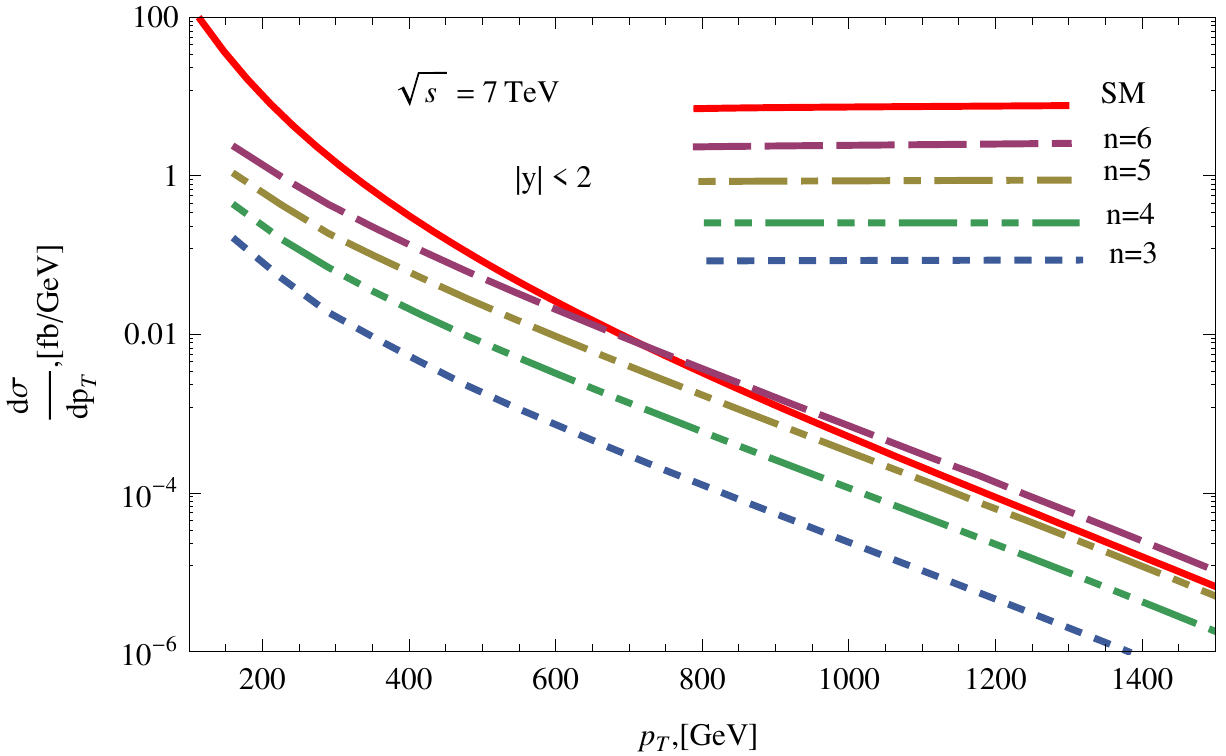}
\end{center}
{\caption \qquad{\em Same as in Fig.~\ref{figure1}, but for
the center-of-mass energy of incoming protons equal to 7~TeV.  }
 \label{figure7} }
\end{figure} 
Due to the relations (\ref{cross_sym1}) and (\ref{cross_sym2}),
 only one squared amplitude for the subprocesses   needs to be calculated. For the bulk $Z$-boson in the final state we obtain
\begin{equation}
\overline{\sum}|\mathcal{M}(\bar{q}q\to gZ_{bulk})|^2 = 8\pi^2 \alpha \alpha_S
(N^2_c-1) \sum_{\lambda = L,R}\left( I_{q_\lambda} \right)^2 
\frac{\hat{t}^2+\hat{u}^2+2m^2 \hat{s}^2}{\hat{t} \hat{u}} 
\cdot \frac{2}{n k}\, Z^2_m(0), \label{amplitude_Zbulk}
\end{equation}
where $\alpha=e^2/(4\pi)$ and $\alpha_S=g^2_S/(4\pi)$ are the 
electromagnetic and strong couplings, and $Z_m(0)$ is the  
wave function of the bulk $Z$-boson given by Eq.~(\ref{Z_m(0)tot}). 
The factor  $I^Z_{q_\lambda}$ is the combination of the weak 
isospin $T_{q_\lambda}^3$ and weak hypercharge $Y_{q_\lambda}$:
$$
I^Z_{q_\lambda}=T^3_{q_\lambda} \,\frac{\cos\theta_W}{\sin\theta_W}
-\frac{Y_{q_\lambda}}{2}\,\frac{\sin\theta_W}{\cos\theta_W}. 
$$ 
The amplitude similiar to (\ref{amplitude_Zbulk}) with $\gamma_{bulk}$ in the final state reads
\begin{equation}
\overline{\sum}|\mathcal{M}(\bar{q}q\to g\gamma_{bulk})|^2 = 16\pi^2 \alpha \alpha_S
(N^2_c-1)Q^2_q
\frac{\hat{t}^2+\hat{u}^2+2m^2 \hat{s}^2}{\hat{t} \hat{u}} 
\cdot \frac{2}{n k}\, A^2_m(0), \label{amplitude_gamma_bulk}
\end{equation}
where $Q_u=2/3$, $Q_d=-1/3$ are the quark electric charges, and
the factor $A_m(0)$ is given by Eq.  (\ref{photon_WF_at_zero}).

\section{ Signal at the LHC \label{section7}} 

In this section the distributions in jet transverse momentum
and jet rapidity are calculated for the process 
$pp  \to \mbox{jet} +\not\!\!E_{T}$, where the energy is carried away 
from  the brane by either bulk 
$Z$-boson or bulk photon.
 We compare these distributions with the main
Standard Model background that comes from the processes 
$pp\to \mbox{jet}+\nu\bar{\nu}$. This background has been computed using the
program  {\tt COMPHEP} \cite{Boos:2009un}.  In our numerical 
calculations, GRV LO PDFs \cite{Gluck:1998xa} are used throughout. The 
factorization scale
of the PDFs is fixed at $\mu=1$ TeV. Only $u$ and $d$ flavors are 
activated since numerical calculations show that the contributions
of the other flavors are negligible.
\begin{figure}[t]
\begin{center}
\includegraphics[width=0.9\textwidth]{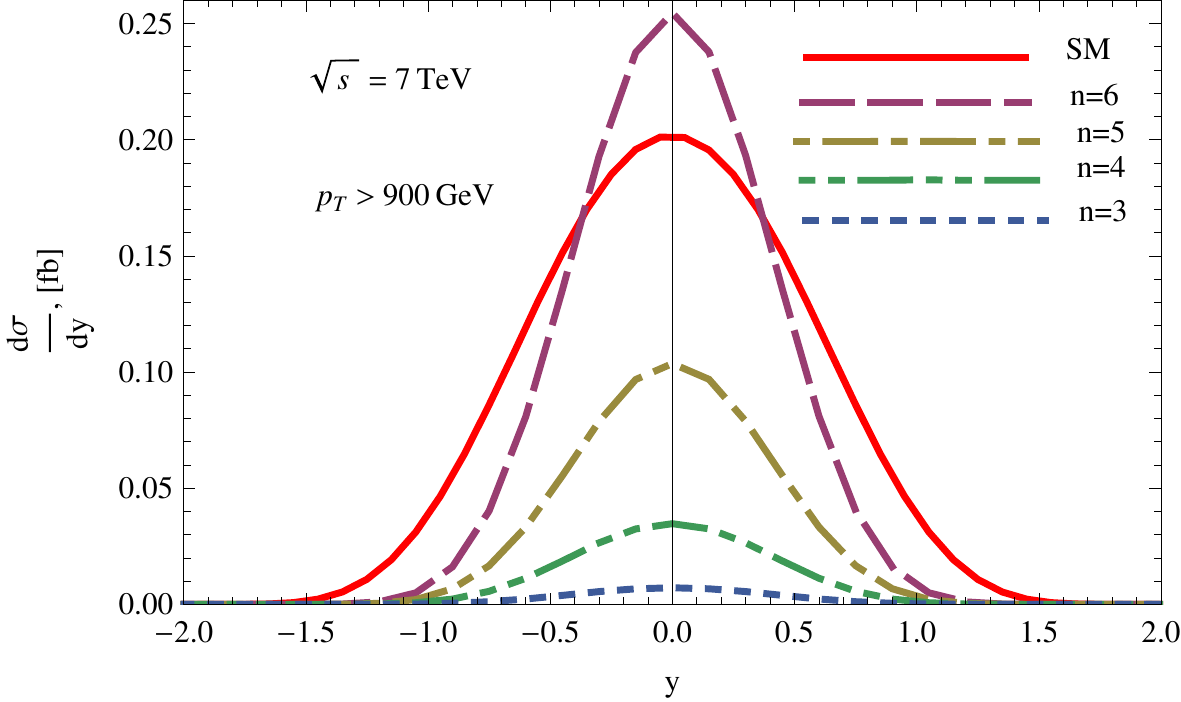}
\end{center}
{\caption \qquad{\em  Same as in Fig.~\ref{figure5}, but for the 
center-of-mass energy of incoming protons equal to 7~TeV.}
 \label{figure6} }
\end{figure}   
We begin our  discussion with the case of the proton collision 
energy  equal to 14 TeV. In Fig. \ref{figure1}  we show $p_T$  
distributions  for the processes $pp\to \mbox{jet}+Z_{bulk}(\gamma_{bulk})$ 
for various numbers of compact extra dimensions $n=3,4,5,6$. The cut 
on the jet  rapidity is $|y| < 2$ for both signal  and background.  
For $n=6$ and $n=5$, the signal  cross section   dominates 
over the Standard 
Model for $p_T> 500$ GeV   and  $p_T> 750$  GeV, respectively. The 
signal is below the background for $n=3,4$. It is clear from Fig. 
\ref{figure1} that   the   cross 
section of $pp\to \mbox{jet}+Z_{bulk}(\gamma_{bulk})$  grows with the increase
of  $n$. This is mainly because larger values of $k$ are allowed
for larger $n$, see Table \ref{table1}. We pointed out in section 
\ref{section5}
( see also Figs.~\ref{figure2},~\ref{figure3}) that the 
effective coupling of bulk fields
has a plateau in the high region. This explains the fact that
 for any given  number of 
 compact dimensions, the ratio of signal to background  is higher in 
 the  high mode region, and hence at larger $p_T$.
 The jet rapidity distributions with the cut $ p_T > 300$ GeV 
are shown in Fig. \ref{figure4}. These distributions are 
correlated with the plots shown in Fig. \ref{figure1}, since the  
main contribution to the SM background comes from the region
$300 \mbox{ GeV} < p_T< 750 \mbox{ GeV} $. To enhance the signal 
with respect to background we consider also the cut
$p_T>900 \mbox{ GeV}$.  This is shown in 
Fig. \ref{figure5}. In the cases  $n=5,6$ signal cross 
sections are larger than the Standard Model background; for
$n=4$ the signal is not negligibly small either. 
These results are  in agreement with the $p_T$ distributions shown 
in  Fig.  \ref{figure1}.

Now we turn to case $\sqrt{s}=7$ TeV. In Fig. \ref{figure7} 
we show the jet transverse
momentum distribution with the cut  $|y|<2$. Obviously, the 
situation is worse at this energy, and only the case 
$n\ge 6$ can possibly be probed. This is also clear from Fig. 
\ref{figure6}, where we plot the jet rapidity distribution with the cut
$p > 900$ GeV.

\section{Summary \label{summary}}
In this paper we have performed the study of the production of the 
bulk  $Z$-boson and photon  at the LHC in the framework of RSII-$n$ 
model.  This process would show up as  
$pp\to \mbox{jet}+\not\!\!E_T$.  The differential distributions in 
jet rapidity and jet 
transverse  momentum  have been calculated and compared with the SM 
background.  Our analys shows that at the total energy $14$ TeV, 
models with  the numbers of compact extra dimensions $n\ge4$ can be 
probed, provided  that the values of the  parameter $k$ of these 
models is  close to the  existing lower bounds. The sensivity is much 
worse at  the total  collision energy 7 TeV: in that case, one can at 
best  start probing  the models with $n\ge6$.

\section{Acknowledgements}
We are indebted to E.~E.~Boos, S.~V.~Demidov, S.~N.~Gninenko, 
D.~S.~Gorbunov, A.~L.~Kataev,  M.~Y.~Kuznetsov, D.~G.~Levkov,   
A.~G.~Panin, V.~A.~Rubakov and M.~A.~Smolyakov 
for helpful  discussions.  This  work  was supported in part by 
grants of  Russian Ministry of  Education 
and  Science NS-5590.2012.2 and
GK-16.740.11.0583, grants of the President of 
Russian Federation MK-2757.2012.2 and MK-1632.2011.2.

\end{document}